# Unidirectional propagation of designer surface acoustic waves

Jiuyang Lu,[1] Chunyin Qiu,[1*] Manzhu Ke,[1] and Zhengyou Liu[1,2*]

[1]Key Laboratory of Artificial Micro- and Nano-structures of Ministry of Education and School of Physics and Technology, Wuhan University, Wuhan 430072, China

[2]Institute for Advanced Studies, Wuhan University, Wuhan 430072, China

**Abstract:**

We propose an efficient design route to generate unidirectional propagation of the designer surface acoustic waves. The whole system consists of a periodically corrugated rigid plate combining with a pair of asymmetric narrow slits. The directionality of the structure-induced surface waves stems from the destructive interference between the evanescent waves emitted from the double slits. The theoretical prediction is validated well by simulations and experiments. Promising applications can be anticipated, such as in designing compact acoustic circuits.

*Authors to whom correspondence should be addressed.
Emails: cyqiu@whu.edu.cn; zyliu@ whu.edu.cn

Surface acoustic waves (SAWs), e.g., Rayleigh and Stoneley surface waves, have been widely involved in electronics, microfluidics and geophysics [1]. Recently, as a counterpart of the structure-induced optic surface waves [2,3], the designer SAW (DSAW) has attracted much attention [4-8]. Different from the conventional SAWs sustained on a planar interface, the DSAW is produced by a periodically textured solid surface placed in a fluid background. The properties of such DSAWs (e.g., dispersion relations) depend mostly on the geometry of the surface corrugations, but less on the acoustic parameters of the constituent ingredients [8]. This leads to great flexibility in engineering the DSAWs by artificially modulating the surface structures. Benefited from the constructive coupling with DSAWs, extraordinary transmissions and beam collimations [4-7] of sound waves have been observed. The direction of sound beam can also be steered by employing asymmetric surface gratings [9]. Note that the aforementioned sound manipulations are closely related to the coupling between the bulk wave and DSAW. The DSAW itself can also be flexibly manipulated as well (although less studied). Examples can be referred to the subwavelength imaging and focusing for two-dimensional (2D) DSAWs [10,11], and the rainbow effect for one-dimensional (1D) DSAWs [12].

In this paper, we propose a simple recipe to generate unidirectional propagation of DSAWs. The peculiar DSAW manipulation is realized by an air-surrounded hybrid structure: a plate patterned with 1D periodical corrugations plus a pair of asymmetric narrow slits drilled through. Principally, the asymmetric double slits generate asymmetric spatial Fourier spectra, while the periodical structure picks out specific evanescent components, i.e., the counter-propagating DSAW modes according to the dispersion relation. When the slit geometries are fully-optimized, the weight contrast between the selected DSAWs (traveling along opposite directions) can be enhanced strikingly, giving rise to nearly unidirectional propagation of DSAWs. This design route has been well validated by full-wave simulations and experimental measurements. Moreover, our study indicates that for an elaborately designed sample, the directionality could be flexibly switched by adjusting the frequency of incident sound. Although the asymmetric propagation of *bulk sound waves* has been extensively investigated in recent years [13-18], based on the asymmetric diffraction effect [13-15] or breaking the time-reversal symmetry [16-18], there is no attention paid to the asymmetric or unidirectional manipulations on *surface waves*. Obviously, the simple design introduced here will open new opportunities for SAW manipulations, similar to the prosperous unidirectional steering of surface Plasmons [19-22].

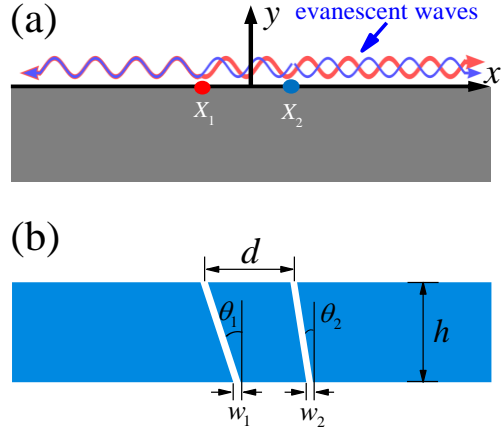

FIG. 1. (Color online) (a) Schematics of the unidirectional propagation of evanescent waves emitted from two idealized line sources (indicated by the red and blue circles), where the right-propagation is prohibited due to the completely destructive interference. (b) A practical realization for (a), where the line sources are mimicked by the sound waves emitted from a pair of asymmetric narrow slits.

To illustrate our idea, as shown in Fig. 1(a), we consider first two ideal line sources placed at ($X_i, 0$), where the subscripts $i = 1, 2$ denote the left and right sources, respectively. The radiation field from the source $i$ can be described as $p_i(\mathbf{r}_i) = A_i e^{i\phi_i} H_0^{(1)}(k_0 |\mathbf{r}_i|)$, where $A_i$ and $\phi_i$ stand for its amplitude and phase, respectively, $H_0^{(1)}$ is the 0-order Hankel function of the first kind, $k_0$ is the wave number of air, and $\mathbf{r}_i$ is a position vector measured from the source $i$. The corresponding Fourier spectrum evaluated at $y = 0$ can be expressed as $F_i(k_x) = \dfrac{2A_i e^{i\phi_i - k_x X_i}}{\sqrt{k_0^2 - k_x^2}}$, where the Fourier components with $|k_x| > k_0$ correspond to evanescent waves (EWs). Assume that the EWs with wavevectors $\pm k_s$ ($k_s > 0$) are desired to be manipulated. A complete suppression of the right-propagating EW requires $F_1(k_s) + F_2(k_s) = 0$, which leads to the amplitude condition $A_1 = A_2$ and the phase condition $(\phi_1 - \phi_2) + k_s(X_2 - X_1) = (2n+1)\pi$ simultaneously, with $n$ being an arbitrary integer. These conditions can be intuitively understood by the destructive interference between the EWs (with $k_x = k_s$) emitted from the two line sources: as schematically manifested in Fig. 1(a), at the right hand side of the both sources, the two right-propagating EWs vibrate with identical amplitudes but opposite directions. Usually, it is also highly desirable (but not necessary) to enhance the left-propagating DSAW, which requires another phase condition of constructive interference, i.e., $(\phi_1 - \phi_2) + (-k_s)(X_2 - X_1) = 2m\pi$. Similar conditions can be also written if the left-propagation is desired to be prohibited.

The functionality of such double line sources can be mimicked by the sound waves emitted from a pair of narrow slits drilled through an epoxy plate (of thickness $h$). As illustrated in Fig. 1(b), the slits have tilt angles $\theta_i$ and widths $w_i$ at the slit exits ($i = 1, 2$). For simplicity, here the system is incident normally by a Gaussian beam from the bottom, associated with identical initial phases at the slit entrances. Considering the great impedance mismatch between air and epoxy, the slotted plate can be modeled as acoustically rigid. The self-consistent sound field can be solved by the coupled-mode theory, in which only the fundamental waveguide mode is allowed inside the narrow slit. Without tedious derivations presented here, the Fourier spectrum for the sound field emitted from the slit $i$ can be evaluated as $F_i^{\text{slit}}(k_x) = \dfrac{2T_i \operatorname{sinc}(k_x w_i/2) e^{-ik_x X_i}}{\sqrt{k_0^2 - k_x^2}}$, where the complex coefficient $T_i$ is determined by the geometries of both slits together, since the slit-slit coupling has been taken into account. For the spatial frequency $k_x w_i \ll 1$ involved here, $\text{sinc}(k_x w_i / 2) \approx 1$ and thus $F_i^{\text{slit}}(k_x)$ is proportional to the quantity $F_i(k_x)$ for ideal line source. This indicates that the sound field radiated from each slit indeed resembles an idealized line source, except for the EWs with much larger wavevectors (which can distinguish the slit width). Therefore, to produce a nearly pure left-propagation of DSAW, one can minimize the ratio $F(+k_s)/F(-k_s)$ by optimizing the slit geometries, where $F(k_x)$ stands for the Fourier spectrum of the total field emitted from the double slits.

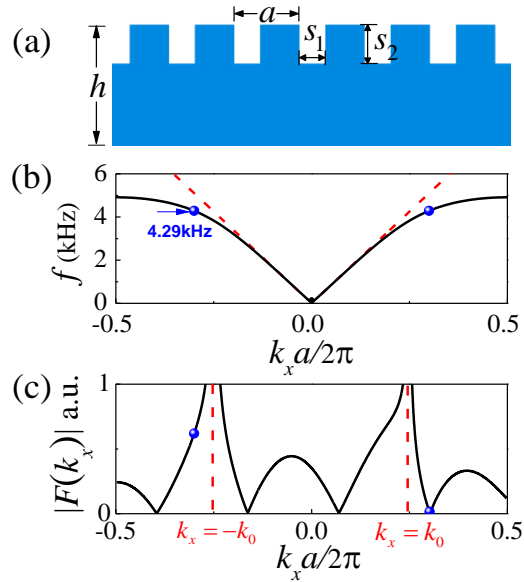

FIG. 2. (Color online) (a) A 1D periodical array of rectangular grooves textured on an epoxy plate. (b) The dispersion relation (black solid) of the DSAW supported by the structure, together with the air line (red dashed) for comparison. (c) The spatial Fourier spectrum for the sound emitted from the double slits with optimized parameters, evaluated at $y = 0$ for 4.29kHz. In (b) and (c), the blue circles show the DSAW modes to be manipulated.

Now we apply the above design to manipulate the DSAW that is supported on an epoxy plate textured with periodical grooves. As depicted in Fig. 2(a), the structure period $a = 2\text{cm}$, the groove width $s_1 = 0.8\text{cm}$, and the groove depth $s_2 = 1.2\text{cm}$. In Fig. 2(b) we present the dispersion curve (solid) for the textured structure, together with the air line for comparison (dashed). Assume that 4.29kHz is the frequency to be manipulated, associated with Bloch wavevectors $k_s = \pm 1.19 k_0$. To suppress the right-propagating DSAW, the ratio $F(+k_s)/F(-k_s)$ is desired to be minimized by exploring the parameter space of slit geometries. To reduce the calculation, only the slit width $w_1$ and the inclined angle $\theta_1$ are searched, which are robust to adjust the amplitude and phase responses in the outgoing facet of the left slit. The remainder parameters are listed as follows: the plate thickness $h = 11.1\text{cm}$, the slit width $w_2 = 0.3\text{cm}$, the inclined angle $\theta_2 = 0$, and the separation $d = X_2 - X_1 = 8.5\text{cm}$. The optimization procedure gives rise to $w_1 = 0.49\text{cm}$ and $\theta_1 = 15.5°$. Based on such set of geometric parameters, the coupled-mode theory gives $T_1/T_2 \approx 1.04$ and $(\phi_1 - \phi_2) + k_s(X_2 - X_1) \approx 0.95\pi$, which approximately satisfy the amplitude and phase conditions for the unidirectional manipulation of DSAW along +x direction. Moreover, the phase difference $(\phi_1 - \phi_2) + (-k_s)(X_2 - X_1) \approx -4.11\pi$ guarantees a constructive interference between the EWs traveling along –x direction. It is worth pointing out that although the slit width $w_2$ is smaller than $w_1$, the amplitude condition holds well since the length of the right slit (*h*) approaches to 1.5 wavelength in free space, which contributes to a strong amplitude response at the slit exit due to the Fabry-Perot resonance. The enhanced amplitude (by using this *h*) together with the constructive interference enables an optimal conversion efficiency to the left-propagating DSAW. In Fig. 2(c) we present the Fourier spectrum $|F(k_x)|$ for the total field evaluated at $y = 0$. Besides the two sharp peeks located at $k_x = \pm k_0$ (as predicted in the idealized model), a particularly important feature is that the Fourier spectrum is asymmetric about $k_x = 0$. Specifically, the ratio $F(+k_x)/F(-k_x)$ reaches a minimum of ~0.02 at the prescribed wavevector of DSAW, i.e., $k_s = 1.19 k_0$.

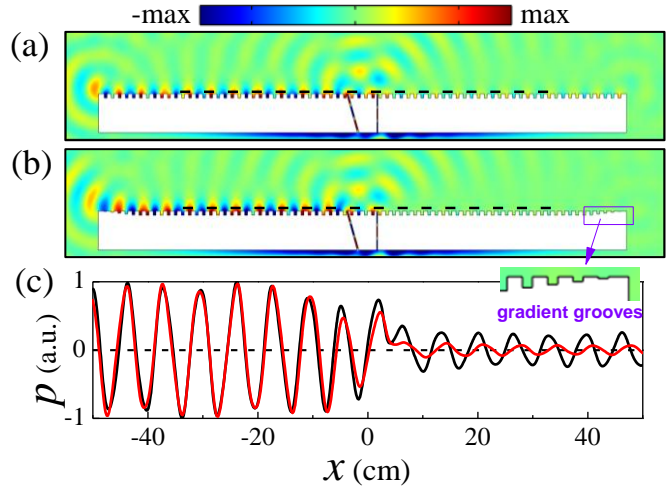

FIG. 3. (Color online) Simulated pressure distributions for the hybrid slit-groove structures (white regions), incident normally by a Gaussian beam from the bottom, where (a) for purely periodical grooves, and (b) for the case with gradient grooves at the lateral sides. (c) The pressure fields distributed along the horizontal dashed lines depicted in (a) and (b), manifested by black and red lines respectively.

To check the validity of the whole design strategy, full-wave simulations for the hybrid slit-groove structure has been carried out based on a finite element solver (COMSOL MULTIPHYSICS package). The whole sample has a length of 150cm, perforated with 74 grooves in total. In Fig. 3(a) we present the pressure field distribution generated by a Gaussian beam incident normally upon the sample. It is observed that, compared with the left-propagation, the right-propagation of DSAW is much weaker. In fact, the right-propagating DSAW is generated mostly from the unwanted reflection of the left-propagating DSAW when approaching the left edge of the sample. To reduce the finite size effect, in each side of the original structure five identical grooves are replaced by the ones with gradually diminished depths. As shown in Fig. 3(b), now the right-propagation of DSAW is indeed reduced considerably. The asymmetric DSAW propagation can be seen more clearly in Fig. 3(c), which displays the pressure field distributed slightly above the sample. Here the black and red lines represent the cases without or with supplying gradient grooves, respectively. [Note: The pressure fields in Fig. 3(c) exhibit an oscillation period of

~6.7cm, which is consistent with the wavelength of the prescribed DSAWs.] To characterize the asymmetric propagation of DSAW, a power distinction ratio $r = \eta_- / \eta_+$ can be defined, where $\eta_-$ and $\eta_+$ correspond to the average pressure intensities in the left- and right-hand sides of the double slits, respectively. In this case, the power distinction ratio is greatly improved from $r = 21$ to 418 when employing anti-reflective grooves, which demonstrates a nearly unidirectional propagation behavior.

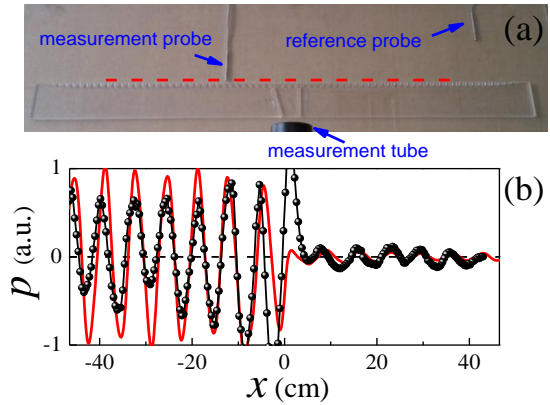

FIG. 4. (Color online) (a) Experimental setup. (b) The pressure field (black dots) measured along the horizontal dashed line in (a), together with the numerical data (red line) for comparison.

Note that our problem is a pure 2D one in *x-y* plane. As shown in Fig. 4(a), in practical experiment the 2D sound propagation is realized in a waveguide of height 1.2cm, which is narrow enough with respect to the wavelength under consideration. The sample is prepared by sculpturing an epoxy plate with double-slits and a 1D periodical array of grooves, together with gradient grooves in lateral as shown in Fig. 3(b). The sample is incident normally by a Gaussian beam launched from the measurement tube. The distribution of the pressure field behind the sample can be scanned by a probe microphone (B&K Type 4187) of diameter 0.7cm, together with another microphone for phase reference. The launched and received sound signals are analyzed by a multi-analyzer system (B&K Type 3560B), from which both of the wave amplitude and phase can be extracted. In Fig. 4(b), we present the pressure distribution for 4.29kHz, measured along a horizontal line with 0.8cm (~0.1 wavelength in air, as mentioned before) away from the sample surface. As predicted, Fig. 4(b) shows a much weaker DSAW propagation along $+x$ direction, comparing with that along $-x$ direction. The power distinction ratio can be roughly evaluated as $r = 38$, which is considerably high despite much smaller than the numerical prediction $r = 418$. This degradation stems mostly from the experimental error in controlling slit widths (note that because of the existence of double slits, the whole sample is assembled by three separate segments).

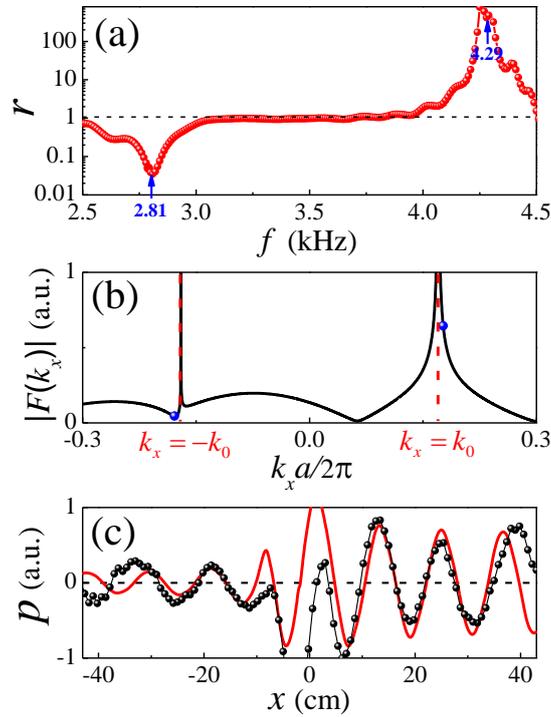

FIG. 5. (Color online) (a) The numerical power distinction ratio $r$ plotted as a function of frequency. (b) The spatial Fourier spectrum for 2.81kHz, where the blue circles indicate the DSAW modes to be manipulated. (c) The pressure distributions measured slightly above the sample, where the black dot and red line correspond to the experimental and numerical results, respectively.

The frequency response of the asymmetric DSAW propagation is of particular interest although the sample design is started from 4.29kHz. In Fig. 5(a) we present the frequency dependent power distinction ratio $r = \eta_- / \eta_+$, extracted numerically from 2.5kHz to 4.5kHz and plotted in a logarithmic scale. It shows a remarkable peak ($r \approx 420$) centered at the prescribed 4.29kHz, around which a highly asymmetric ($r > 20$) DSAW propagation can be observed from 4.21kHz to 4.36kHz. From the frequency 3.0kHz to 4.0kHz, there appears a wide plateau associated with $r \sim 1$, which means a nearly symmetric DSAW propagation in this frequency interval. As the frequency is further reduced, an obvious dip ($r \approx 1/29$) emerges at 2.81kHz, indicating a dominant DSAW propagation along $+x$ direction. To understand this unexpected behavior, in Fig. 5(b) we present the spatial Fourier spectrum for 2.81kHz. It shows a minimum at $k_x = -1.04k_0$, which corresponds exactly to the negative wavevector determined by the dispersion curve. Besides, based on the coupled-mode theory, the wave responses at the outgoing facets give rise to $T_1/T_2 \approx 0.98$ and $(\phi_1 - \phi_2) + (-k_s)(X_2 - X_1) \approx -1.06\pi$, which approximately satisfy the amplitude and phase conditions for destructive interference. Therefore, the structure originally designed for the left-propagating DSAW can support a right-propagating DSAW at another frequency [23]. This property has also been checked experimentally by the pressure distribution measured along the aforementioned horizontal line, as shown by the black dots in Fig. 5(c), together with the numerical result for comparison. The experimental power distinction ratio is $r \approx 1/7$, which degrades with respect to the simulation due to the experiment accuracy again. It is expected that this switch effect (by frequency) can be further improved by exploring a broader parameter space for the double slits.

    Starting from an analytical model, we have introduced an elegant scheme to realize the unidirectional propagation of the DSAW supported on a textured surface. Different from the approaches used in plasmonic systems, which usually employ the polarization of light [19-22], here the highly asymmetric DSAW propagation comes from the destructive interference of the EWs emitted from two separated sound sources. The capability of the switchable directionality (by frequency) has also been discussed. Prospective applications can be anticipated for the unidirectional DSAW propagation, such as in acoustic integrated devices or in ultrasonic detection devices.

**Acknowledgments**

This work is supported by the National Natural Science Foundation of China (Grant Nos. 11174225, 11004155, 11374233, and J1210061); the National Basic Research Program of China (Grant No. 2015CB755500); the Program for New Century Excellent Talents (NCET-11-0398); and the Fundamental Research Funds for the Central Universities (Grant No. 2014202020204).**Acknowledgments**

This work is supported by the National Natural Science Foundation of China (Grant Nos. 11174225, 11004155, 11374233, and J1210061); the National Basic Research Program of China (Grant No. 2015CB755500); the Program for New Century Excellent Talents (NCET-11-0398); and the Fundamental Research Funds for the Central Universities (Grant No. 2014202020204).